\documentclass[sn-mathphys-num]{sn-jnl}% Math and Physical Sciences Numbered Reference Style

\usepackage{graphicx}%
\usepackage{multirow}%
\usepackage{amsmath,amssymb,amsfonts}%
\usepackage{mathrsfs}%
\usepackage[title]{appendix}%
\usepackage{xcolor}%
\usepackage{textcomp}%
\usepackage{manyfoot}%
\usepackage{booktabs}%
\usepackage{algorithm}%
\usepackage{algorithmicx}%
\usepackage{algpseudocode}%
\usepackage{listings}%
\usepackage{doi}%
\newcommand{\rmd}{{\rm d}}
\newcommand{\rmi}{{\rm i}}
\newcommand{\rme}{{\rm e}}

\begin{document}

\title[SUSY Pauli Hamiltonians]{On the SUSY structure of spherically symmetric Pauli Hamiltonians}

\author{\fnm{Georg} \sur{Junker}}\email{georg.junker@fau.de}

\affil{\orgdiv{Inst.\ f\"ur Theoretische Physik I}, \orgname{Friedrich-Alexander-Universit\"at}, \orgaddress{\street{Staudtstr.\ 7 B 3}, \postcode{91058} \city{Erlangen}, \country{Germany.}}}

\affil{~\\ and}

\affil{\orgdiv{}\orgname{~\\ European Southern Observatory}, \orgaddress{\street{Karl-Schwarzschild-Str.\ 2}, \postcode{85748} \city{Garching bei M\"unchen}, \country{Germany.}}}

\abstract{It is shown that the quantum Hamiltonian characterising a non-relativistic electron under the influence of an external spherical symmetric electromagnetic potential exhibits a supersymmetric structure. Both cases, spherical symmetric scalar potentials and spherical symmetric vector potentials are discussed in detail. The current approach, which includes the spin-$\frac{1}{2}$ degree of freedom, provides new insights to known models like the radial harmonic oscillator and the Coulomb problem. We also find a few new exactly solvable models, one of them exhibiting a new mixed type of shape invariance containing translation and scaling of potential parameters. The fundamental role as Witten parity played by the spin-orbit operator is high-lighted.}


%%\pacs[JEL Classification]{D8, H51}

%%\pacs[MSC Classification]{35A01, 65L10, 65L12, 65L20, 65L70}

\maketitle

\section{Introduction}\label{sec1}
Supersymmetric quantum mechanics (SUSY QM) was originally introduced by Nicolai when studying supersymmetry in spin systems \cite{Nic1976,Nic1977}. SUSY QM become popular when Witten \cite{Witten1981,Witten1982} introduced a simple one-dimensional model for SUSY QM. Since then SUSY QM has become a highly appreciated tool in various areas of physics \cite{Cooper1995,Junker1996,Cooper2001,Gangopad2010}. Most of the applications of SUSY QM make either use of Witten's non-relativistic quantum model in one dimension or utilise the relativistic supersymmetric Dirac equation mainly in two dimension when studying properties of graphene. Applications of the SUSY QM formalism in Pauli systems characterising a non-relativistic charged spin-$\frac{1}{2}$ particle in three dimensions are still rare. Here a SUSY structure can be obtained for an electron in an arbitrary magnetic field resulting in a gyromagnetic ration $g=2$, which otherwise may only be derived for Dirac's equation. In general this system does not provide a Witten parity operator. Hence, an essential part of the SUSY structure is missing. Only in the special case of a unidirectional magnetic field, say in the $z$-direction, such an operator exists \cite{Junker1996}. That is, while in two dimensions the Pauli Hamiltonian for a spin-$\frac{1}{2}$ particle naturally incorporates SUSY as the spin component orthogonal to the plane, represented by the Pauli matrix $\sigma_z$, can act as the Witten parity operator, in three dimensions SUSY Pauli systems are rather sparse and limited to special forms of the external magnetic field like that a magnetic monopole \cite{DHoker1084} or with an inversion invariance \cite{Tkachuk1997}.

The aim of the present work is to enlarge the list of supersymmetric Pauli systems by considering the Pauli Hamiltonian for general spherically symmetric scalar or vector potentials. It is shown that restricting the quantum model onto a subspace of fixed total angular momentum naturally provides a SUSY structure with the spin-orbit operator acting as the Witten parity operator. Various well-known as well as new SUSY models are found by this novel approach.

This paper is organised as follows. In the next section we briefly review the basic of SUSY QM as utilised in our way forward. Section 3 discusses the classes of spherically symmetric models being considered and introduces the subspaces of fixed total angular momentum. We are then able to show, in section 4, that such subspaces may hold a SUSY structure with a build-in Witten parity operator represented by the spin-orbit operator and a generic supercharge, which in essence is defined via a suitable vector operator. Section 5 discusses the case of spherically symmetric scalar potentials, while section 6 deals with the spherically symmetric vector potentials. Both sections present a general discussion as well as various explicit examples. We conclude with a summary and outlook in section 7.

\section{Supersymmetric quantum mechanics}\label{sec2}
A quantum mechanical system is characterised by a self-adjoint Hamiltonian $H=H^\dag$ generating its time evolution. That is, it acts on states in a given Hilbert space ${\cal H}$. If, in addition, there exists a so-called supercharge $Q=Q^\dag$ obeying the relation $H=Q^2$, and a Witten operator $W=W^\dag$ being a unitary involution, $W^2=1$, and anti-commuting with $Q$, $\{W,Q\}:=WQ+QW$, then such a quantum system is supersymmetric. The corresponding SUSY algebra reads
\begin{equation}\label{SUSYALgebra}
  H=Q^2\,,\quad W^2=1\,,\quad \{W,Q\}=0\,,\quad [H,W]=0\,.
\end{equation}
In the last relation above the square brackets denotes the commutator $[H,W]:=HW-WH$ indicating that $H$ and $W$ have a joint set of common eigenstates. Being a self-adjoint unitary involution, the Witten operator has the two eigenvalues given by $\pm 1$, i.e.\ ${\rm spec\,} W=\{-1,+1\}$. This induces a grading of the Hilbert space into its subspaces ${\cal H}_\pm:= \{|\psi\rangle_\pm\in{\cal H}\, |\, W |\psi\rangle_\pm=\pm |\psi\rangle_\pm\}$ of states with positive and negative Witten parity. As $H=Q^2\geq 0$ there exit no negative energy eigenstates. When zero-energy eigenstates exit then SUSY is said to be unbroken. In the case $H>0$ SUSY is said to be dynamically broken. Obviously the anti-commutation relation in \eqref{SUSYALgebra} implies $Q: {\cal H}_\pm\to {\cal H}_\mp$. As a direct consequence all positive-energy eigenstates are pairwise degenerate and the supercharge generates the SUSY transformation
\begin{equation}\label{SUSYTrafo}
  Q|\psi_E\rangle_\pm = \sqrt{E} |\psi\rangle_\mp\,, \quad|\psi_E\rangle_\pm\in{\cal H}_\pm\,,\quad\mbox{where}\quad H|\psi_E\rangle_\pm = E |\psi_E\rangle_\pm \quad\forall\quad E>0\,.
\end{equation}

With the help of $Q$ and $W$ one may construct a pair of self-adjoint supercharges $Q_1:=Q$ and $Q_2:=-\rmi WQ=\rmi QW$ obeying the alternative SUSY algebra
\begin{equation}\label{SUSYAlgbera2}
  \{Q_i,Q_j\} = 2H\delta_{ij}\,,\qquad i,j=1,2\,.
\end{equation}
From these SUSY charges the Witten parity may be reconstructed, $W:=\frac{\rmi}{H}\,[Q_1,Q_2]$, obeying $\{Q_i,W\}=0$.
Another realisation of SUSY QM is obtained by introducing a  complex supercharge, $Q_c:=\frac{1}{\sqrt{2}}\,\{Q_1+\rmi Q_2\}$ obey the algebra
\begin{equation}\label{SUSYAlgebra3}
  H=\{Q_c,Q_c^\dag\}\,,\qquad Q_c^2=0\,.
\end{equation}
Again the Witten parity can be reconstructed via $W:=\frac{[Q_c,Q_c^\dag]}{\{Q_c,Q_c^\dag\}}$ with $\{W,Q_c\}=0$. All these three realisations of SUSY QM are equivalent and used in the literature. As a final comment let us mention that the reconstruction of $W$ must be done in the subspace spanned by the positive energy eigenstates with $E>0$ and then naturally extended onto the full Hilbert space in the case of unbroken SUSY.
\section{Pauli Hamiltonians with spherical symmetry}\label{sec3}
The quantum dynamics of a non-relativistic spin-$\frac{1}{2}$ particle with mass $m>0$ and charge $e$, with $e<0$ for an electron, subjected to an external electromagnetic field characterised by scalar potential $\phi$ and vector potential $\vec{A}$, is represented the Pauli Hamiltonian
\begin{equation}\label{PauliH}
  H_P:= \frac{1}{2m}\left(\vec{P}-\frac{e}{c}\vec{A}(\vec{r})\right)^2-\frac{e}{2mc}\vec{B}(\vec{r})\cdot \vec{\sigma}+e\phi(\vec{r})\,.
\end{equation}
Here we have set the gyromagnetic ratio to $g=2$, Planck's constant $\hbar =1$, $c$ stands for the speed of light, the magnetic field is given by $\vec{B}(\vec{r})=\vec{\nabla}\times\vec{A}(\vec{r})$ and $\vec{\sigma}/2$ represents the spin operator for a spin-$\frac{1}{2}$ degree of freedom via Pauli matrices. Obviously, the corresponding Hilbert space is ${\cal H}=L^2(\mathbb{R}^3)\otimes\mathbb{C}^2$.

Restricting ourselves to spherical symmetric systems, we consider scalar potentials depending only on $r=|\vec{r}|$, that is, being of the form $V(r) := e\phi(\vec{r})$, where we absorb the electric charge in $V$. Similar, for the vector potential we assume a radial dependency of the form $\vec{A}(\vec{R})=\vec{a}(r)\vec{e}_r$, where $\vec{e}_r$ denotes the unit vector in radial direction. It is easy to see that such vector potentials result in a vanishing magnetic field due to the non-existing of magnetic monopoles. Hence, the vector potential can be represented by a gradient field. This would be a trivial ingredient as long as the associated gauge field is real. Non-trivial physics could only be accomplished when we allow for a pure complex gauge field as in the case of the well-studied Dirac oscillator \cite{Ito,Moshinsky}. Hence, in going forward we will consider the two cases
$$
\begin{array}{llll}
 Case~ I:\quad&  V(r)=e\phi(r)\,,\quad&\vec{A}=\vec{0}\,,\qquad&\mbox{pure~scalar~potential},\\[2mm]
 Case~ II:\quad& V(r)=0\,,\quad&\vec{A}=\rmi\vec{\nabla}U(r)\,,\qquad&\mbox{pure~vector~potential.}
\end{array}$$
For both cases we have a spherical symmetry implying below conserved, not necessarily independent, observables.\\

\begin{tabular}{lll}
  %\hline
  % after \\: \hline or \cline{col1-col2} \cline{col3-col4} ...
   {\bf Operator} & {\bf Definition} & {\bf Quantum Number} \\[2mm] %\hline
   orbit angular momentum\qquad~ & $\vec{L}:= \vec{r}\times\vec{P}$\qquad ~& $\ell=0,1,2,3,\ldots$ \\[1mm]
   spin angular momentum  & $\vec{S}:=\frac{1}{2}\,\vec{\sigma}$& $m_s=\pm\frac{1}{2}$ \\[1mm]
   total angular momentum  & $\vec{J}:=\vec{L}+\vec{S}$ & $j=\frac{1}{2},\frac{3}{2},\frac{5}{2},\ldots$ \\[1mm]
                           & $J_z := L_z+S_z$ & $m_j=-j,\ldots, +j$\\[1mm]
   spin-orbit operator&     $K:= 2\vec{S}\cdot\vec{L}+1$& $\kappa=-s(j+\frac{1}{2})\,, s=\pm 1$
  %\hline
\end{tabular}\\

\noindent
Let us choose as independent quantum numbers the set $(j,m_j,s)$, where $s=\pm 1$ indicates the sign of the eigenvalue of $K$. As a consequence, when keeping $j$ fixed, the angular momentum quantum number becomes $s$-dependent
\begin{equation}\label{ls}
  \ell(s)=j-\frac{s}{2}\,.
\end{equation}
The corresponding angular decomposition of the Hilbert space
\begin{equation}\label{HDecompose}
  {\cal H}= L^2(\mathbb{R}^+,r^2\rmd r)\otimes L^2(S^2)\otimes\mathbb{C}^2
\end{equation}
is then given by
\begin{equation}\label{HDecomp}
  L^2(S^2)\otimes\mathbb{C}^2 = \bigoplus_{j=\frac{1}{2}}^\infty\,\bigoplus_{m_j=-j}^j\,\bigoplus_{s=\pm 1} |j,m_j,s\rangle\langle j,m_j,s|\,.
\end{equation}
Here $|j,m_j,s\rangle$ denotes the so-called Pauli spinor, a two-spinor, which in coordinate representation is given via the usual spherical harmonics $Y_\ell^{m_\ell}$ as follows
\begin{equation}\label{PauliSpinor}
  {\cal Y}^{(s)}_{jm_j}(\theta,\varphi):= \langle\theta,\varphi|j,m_j,s\rangle=
\left(
\begin{array}{c}
  \sqrt{\frac{\ell(s)+sm_j+\frac{1}{2}}{2{\ell(s)}+1}}\, { Y}_{\ell(s)}^{m_j-\frac{1}{2}} \\[2mm]
  s\sqrt{\frac{\ell(s)-sm_j+\frac{1}{2}}{2{\ell(s)}+1}}\, {Y}_{\ell(s)}^{m_j+\frac{1}{2}}
\end{array}
\right).
\end{equation}
This representation induces the partial wave expansion
\begin{equation}\label{PartialWavePsi}
  \Psi(r,\theta,\varphi)=\sum_{j=\frac{1}{2}}^{\infty}\sum_{s=\pm 1}R_{\ell (s)}(r)\sum_{m_j=-j}^{j}{\cal Y}^{(s)}_{jm_j}(\theta,\varphi)\,.
\end{equation}
Let us keep the set $(j,m_j)$ fixed, i.e.\ we limit our discussion to a subspace ${\cal H}_{j,m_j}$, where
\begin{equation}\label{PartialWave}
  {\cal H}=  \bigoplus_{j=\frac{1}{2}}^\infty\,\bigoplus_{m_j=-j}^j{\cal H}_{j,m_j}\qquad\mbox{with}\qquad {\cal H}_{j,m_j}:= L^2(\mathbb{R}^+,r^2\rmd r)\otimes \mathbb{C}^2\,.
\end{equation}
Within such a subspace the radial wave functions may then represented by two-spinors of the form
\begin{equation}\label{radialwave}
  \psi(r)=\sum_{s=\pm1}R_{\ell(s)}\chi^{(s)}=\left(\begin{array}{c}R_{\ell-1}(r)\\ R_{\ell}(r)
  \end{array}\right).
\end{equation}
In the above we have set $\ell :=\ell(-1)$ and as $j$ is fixed we have $\ell(+1)=\ell -1$ according to \eqref{ls}. In other words, the two subspaces corresponding to $s=\pm 1$ are subspaces with orbital angular momentum $\ell$ and $\ell - 1$, respectively. Obviously we need to restrict $\ell =1,2,3,\ldots$ excluding $\ell =0$. The case with vanishing orbital angular momentum is covered by $s=+1$ with $\ell = 1$. Note that the Witten model for the radial harmonic oscillator as well as the radial Coulomb problem comprise a SUSY pair of quantum systems with the same relations between their orbital angular momenta. This is already a first signature of a hidden SUSY. We will uncover this hidden SUSY in the next section.

\section{The hidden SUSY structure}\label{sec4}
As indicated above, their appears to exist a SUSY transformation between the two subspaces ${\cal H}_{j,m_j}$ corresponding to the two possible value of quantum number $s$, which is basically the sign of the eigenvalue of the spin-orbit operator. Hence, we may define a Witten parity operator by
\begin{equation}\label{DefW}
  W:=-\frac{K}{|K|}\,.
\end{equation}
This is a well-defined quantity as $K\neq 0$ and obeys the relation $W^2=1$ as required by SUSY, cf.\ \eqref{SUSYALgebra}. Restricting this operator onto the subspace ${\cal H}_{j,m_j}$ with a fixed value of the total angular momentum $j$ we arrive at the matrix representations
\begin{equation}\label{Krestr}
  \left.K\right|_{{\cal H}_{j,m_j}}=\left(\begin{array}{cc} \ell & 0 \\ 0 & -\ell \end{array} \right)\qquad\mbox{or}\qquad
  K\chi^{(s)}=s \ell \,\chi^{(s)}\,,
\end{equation}
\begin{equation}\label{Wrestr}
  \left.W\right|_{{\cal H}_{j,m_j}}=\left(\begin{array}{cc} 1 & 0 \\ 0 & -1 \end{array} \right)\qquad\mbox{or}\qquad
  W\chi^{(s)}=s\chi^{(s)}\,.
\end{equation}
In addition, we recall \cite{Bjorken1964,Thaller1992} that the operator $\left(\vec{\sigma}\cdot\vec{e}_r\right)$ transforms eigenstates with eigenvalue $s$ into eigenstates with $-s$, that is,
\begin{equation}\label{transop}
  \left.\left(\vec{\sigma}\cdot\vec{e}_r\right)\right|_{{\cal H}_{j,m_j}}=\left(\begin{array}{cc} 0 & 1 \\ 1 & 0 \end{array} \right) \qquad\mbox{or}\qquad
  \left(\vec{\sigma}\cdot\vec{e}_r\right)\chi^{(s)}=\chi^{(-s)}
\end{equation}
That implies the anti-commutation relation $\{ \left(\vec{\sigma}\cdot\vec{e}_r\right), W\}=0$. More general, for an arbitrary radial vector operator of the form $\vec{v}:= v(P_r,K,r)\vec{e}_r$, where
$P_r:=-\rmi \frac{1}{r}\partial_r\, r$ is the self-adjoint radial momentum operator on $L^2(\mathbb{R}^+,r^2\rmd r)$, we have the relations $[v,K]=0$ and
\begin{equation}\label{transop2}
  \{\left(\vec{\sigma}\cdot\vec{v}\right),W\}=0\,.
\end{equation}
With a proper choice of $\vec{v}$ we may then define a self-adjoint supercharge and Hamiltonian via
\begin{equation}\label{SUSYradial}
  Q:=\left(\vec{\sigma}\cdot\vec{v}\right)\,,\qquad H:=Q^2= \left(\vec{\sigma}\cdot\vec{v}\right)^2
\end{equation}
obeying together with $W$ as defined in \eqref{DefW} the SUSY algebra \eqref{SUSYALgebra}. For $H$ representing a non-relativistic Hamiltonian it is obvious that $\vec{v}$ should be in essence linear in the momentum $\vec{P}$.

\section{Case I: Spherically symmetric scalar potentials}\label{sec5}
In this section let us discuss the Case I, i.e., we set $V(r)=e\phi(r)$ and $\vec{A}=\vec{0}$ in the Pauli Hamiltonian \eqref{PauliH} and arrive at
\begin{equation}\label{HPcase1}
  H_{\rm P}=\frac{P^2_r}{2m}+\frac{\vec{L}^2}{2mr^2}+V(r)\,.
\end{equation}
Restricting this operator to the subspace with fixed $j$ we have
\begin{equation}\label{HPcase1restr}
  \left.H_{\rm P}\right|_{{\cal H}_{j,m_j}}=
\left(\begin{array}{cc}H_+ & 0 \\ 0 & H_- \end{array} \right)\qquad\mbox{with}\qquad
  H_\pm = \frac{P^2_r}{2m}+\frac{\ell(\ell\mp 1)}{2mr^2}+V(r)\,.
\end{equation}
Note that $H_\pm$ is not yet a pair of SUSY partner Hamiltonians. We still need to define a proper vector $v$ which in turn will specify the scalar potential $V$.

\subsection{Example 1: The free particle with spin}\label{subsec51}
For the free particle with spin the obvious choice for the vector $\vec{v}$ is
\begin{equation}\label{vfree}
  \vec{v}=\frac{1}{\sqrt{2m}}\,\vec{P}
\end{equation}
resulting in a supercharge of the form
\begin{equation}\label{Qfree}
Q=\frac{\vec{\sigma}\cdot\vec{P}}{\sqrt{2m}}=\frac{\vec{\sigma}\cdot\vec{e}_r}{\sqrt{2m}}\left[P_r +\rmi\frac{K}{r}\right]
=\frac{-\rmi}{\sqrt{2m}}\left(\vec{\sigma}\cdot\vec{e}_r\right)\left[\partial_r -\frac{K-1}{r}\right]
\end{equation}
Restricted onto ${\cal H}_{j,m_j}$ the supercharge reads
\begin{equation}\label{Qfreerestr}
\left.Q\right|_{{\cal H}_{j,m_j}}=\frac{1}{\sqrt{2m}}\left(
\begin{array}{cc}
  0 & P_r -\rmi \frac{\ell}{r} \\
  P_r +\rmi \frac{\ell}{r} & 0
\end{array}
\right)=\frac{-\rmi}{\sqrt{2m}}\left(
\begin{array}{cc}
  0 & \partial_r +\frac{\ell +1}{r} \\
  \partial_r -\frac{\ell -1}{r} & 0
\end{array}
\right)
\end{equation}
and we arrive at the SUSY partner Hamiltonian
\begin{equation}\label{HpmFree}
  H_\pm = \frac{P^2_r}{2m}+\frac{\ell(\ell \mp 1)}{2mr^2}\,.
\end{equation}
The eigenfunctions of these partner Hamiltonians for energy $E=\frac{k^2}{2m}$, $k>0$, are given by normalised spherical Bessel functions $R_\ell(r)= \sqrt{\frac{2k^2}{\pi}}\,j_\ell(kr)$ and explicitly read
\begin{equation}\label{psiFree}
  \psi^+(r)=R_{\ell-1}(r)= \sqrt{\frac{2k^2}{\pi}}\, j_{\ell -1}(kr)\,,\qquad
  \psi^-(r)=R_{\ell}(r)= \sqrt{\frac{2k^2}{\pi}}\,j_{\ell}(kr)\,.
\end{equation}
With the help of the recursion relations for spherical Bessel functions,
\begin{equation}\label{recurBessel}
  \left(\partial_z +\frac{\ell+1}{z}\right)j_\ell(z) = j_{\ell-1}(z)\,,\qquad
  \left(\partial_z -\frac{\ell-1}{z}\right)j_{\ell-1}(z) = j_{\ell}(z)\,,
\end{equation}
one may easily verify the expected SUSY transformation relations
\begin{equation}\label{SUSYTransFree}
  Q\left(\begin{array}{c} \psi^+ \\ \psi^- \end{array}\right) = -\rmi\sqrt{E}\left(\begin{array}{c} \psi^- \\ \psi^+ \end{array}\right)\,.
\end{equation}

\subsection{Example 2: The Coulomb problem with spin}\label{subsec52}
Having discussed the free particle case as a first simple example, let us now turn to the Coulomb problem for a spin-$\frac{1}{2}$ electron orbiting a positive central charge $Ze$. The corresponding Pauli Hamiltonian  then reads
\begin{equation}\label{VCoulomb}
  H_{\rm P} := \frac{\vec{P}^2}{2m}-\frac{\alpha}{r}\,,\qquad \alpha := Ze^2 > 0\,.
\end{equation}
Hence, we need to find a suitable vector $\vec{v}$, which will result in above scalar interactions. As pointed out before such a vector should be linear in the momentum operator. Hence, let us look at the dimensionless version of the Laplace-Runge-Lenz vector
\begin{equation}\label{LRLvector}
  \vec{R}:=\frac{1}{2\alpha m}(\vec{P}\times\vec{L}-\vec{L}\times\vec{P})-\vec{e}_r
\end{equation}
and define
\begin{equation}\label{VQCoulomb}
  \vec{v} := \sqrt{\frac{m\alpha^2}{2K^2}}\,\vec{R}\,,\qquad Q:=\sqrt{\frac{m\alpha^2}{2K^2}}\,\left(\vec{\sigma}\cdot\vec{R}\right)\,.
\end{equation}
A little calculation shows that the SUSY Hamiltonian is up to an additive term identical to \eqref{VCoulomb},
\begin{equation}\label{HSUSYCoulomb}
  H= Q^2 = H_{\rm P} + \frac{m\alpha^2}{2K^2} = \frac{P^2_r}{2m}+\frac{K(K-1)}{2mr^2} -\frac{\alpha}{r}+ \frac{m\alpha^2}{2K^2}\,.
\end{equation}
Hence, the SUSY partner Hamiltonians, when restricting \eqref{HSUSYCoulomb} onto ${\cal H}_{j,m_j}$, are given by
\begin{equation}\label{HpmCoulomb}
   H_\pm = \frac{P^2_r}{2m}+\frac{\ell(\ell \mp 1)}{2mr^2} -\frac{\alpha}{r}+ \frac{m\alpha^2}{2\ell^2}\,,
\end{equation}
which may be identified with the well-known Witten model of the radial Coulomb problem \cite{Junker1996}. In fact, we may be more explicit by noting that
\begin{equation}\label{QCoulomb}
  \displaystyle
 \left(\vec{\sigma}\cdot\vec{R}\right)=\frac{-\rmi}{\alpha m}\left(\vec{\sigma}\cdot\vec{P}\right)K-\left(\vec{\sigma}\cdot\vec{e}_r\right)\,,\qquad
 \displaystyle
 Q=\frac{1}{\sqrt{2m}}\left[\rmi\left(\vec{\sigma}\cdot\vec{P}\right)W-\frac{m\alpha}{|K|}\left(\vec{\sigma}\cdot\vec{e}_r\right) \right].
\end{equation}
Note that for $\alpha=0$ above result in essence reduces to the alternative choice $Q_2=\rmi Q_1W$ of the free partible case discussed before. Restricting the supercharge on the subspace with fixed total angular momentum we arrive at the representation
\begin{equation}\label{QCoulombrest}
  \left.Q\right|_{{\cal H}_{j,m_j}}=\frac{1}{\sqrt{2m}}\left(
\begin{array}{cc}
  0 & \displaystyle \rmi P_r + \frac{\ell}{r} - \frac{m\alpha}{\ell}\\
  \displaystyle -\rmi P_r + \frac{\ell}{r}- \frac{m\alpha}{\ell} & 0
\end{array}
\right)\,,
\end{equation}
which indeed coincides with the Witten model for the radial Coulomb problem \cite{Junker1996}.

\section{Case II: Spherically symmetric vector potentials}\label{sec6}
This section will discuss the second case with a vanishing scalar potential and a non-vanishing radial symmetric vector potential. That is, we assume a vector potential of the form $\vec{A}(\vec{r})=A(r)\vec{e}_r$. As already mentioned in section \ref{sec3}, such a vector potential results in a vanishing magnetic field $\vec{B}=\vec{\nabla}\times\vec{A}=\vec{0}$ and  would, for a real-valued function $A:\mathbb{R}^+\to\mathbb{R}$, simply result in the free particle case discussed in the previous section. For that reason, we consider a vector potential of the form
\begin{equation}\label{a}
  \vec{A}(r):=-\rmi\frac{c}{e} \vec{\nabla}U(r) = -\rmi U'(r)\vec{e}_r\,.
\end{equation}
As we will see below, the function $U:\mathbb{R}^+\to\mathbb{R}$ represents the so-called superpotential \cite{Junker1996}. The idea of such a purely imaginary vector potential is not new and in essence goes back to the so-called Dirac oscillator, where $U(r)=\frac{m}{2}\omega r^2$, originally introduced by It\^{o}, Mori and Carriere \cite{Ito} and discussed in more detail by Moshinsky and Szczepaniak \cite{Moshinsky}. Hence, for defining a proper supercharge we will consider the vector operator via the non-minimal substitution
\begin{equation}\label{vU}
  \vec{P}\quad\to\quad\vec{v} :=  \vec{P}+\rmi U'(r)\vec{e}_r
\end{equation}
and define the now complex supercharge following \eqref{SUSYradial}
\begin{equation}\label{Qc}
  Q_c:= \frac{1}{\sqrt{2m}}\left(\vec{\sigma}\cdot\vec{v}\right)\neq Q_c^\dag\,.
\end{equation}
Here we note that  $Q_c^2\neq 0$, that is, it does not fulfill the nilpotent requirement in \eqref{SUSYAlgebra3}. A way out of this dilemma was pointed out by Ui \cite{Ui} by doubling the Hilbert space in analogy to Dirac's theory. That is, we define the self-adjoint supercharge
\begin{equation}\label{D}
  D:= \left(\begin{array}{cc}0 & Q_c \\ Q_c^\dag & 0 \end{array}\right)
\end{equation}
acting on
\begin{equation}\label{Hdouble}
  {\cal H} := {\cal H}^{(+)}\oplus{\cal H}^{(-)}\,,\qquad {\cal H}^{(\pm)}=L^2(\mathbb{R}^3)\otimes\mathbb{C}^2
\end{equation}
as follows
\begin{equation}\label{DcalH}
 D:{\cal H}^{(\pm)}\to {\cal H}^{(\mp)}\,.
\end{equation}
The corresponding SUSY Hamiltonian is then given by
\begin{equation}\label{HSUSYD}
  H=D^2= \left(\begin{array}{cc}H_{\rm P}^{(+)} & 0 \\ 0 & H_{\rm P}^{(-)} \end{array}\right),\qquad
  H_{\rm P}^{(+)}:= Q_cQ_c^\dag\,,\quad H_{\rm P}^{(-)}:= Q_c^\dag Q_c\,.
\end{equation}
Here the partner Pauli Hamiltonians explicitly read
\begin{equation}\label{HpmD}
  {H}_{\rm P}^{(\pm)}=\frac{\vec{P}^2}{2m}+\frac{{U'}^2(r)}{2m}\mp\frac{U''(r)}{2m}\mp\frac{K}{mr}\,U'(r)
\end{equation}
and act on states in ${\cal H}^{(\pm)}$. We note the additional spin-orbit coupling term appearing as last expression in \eqref{HpmD}.
Similarly the Witten parity is extended to ${\cal H}$ and reads in matrix representation
\begin{equation}\label{WD}
W= \left(\begin{array}{cc}W^{(+)} & 0 \\ 0 & W^{(-)} \end{array}\right) =
\left(\begin{array}{rr|rr}1 & 0 & 0 & 0 \\ 0 & -1 & 0 & 0 \\ \hline 0& 0& 1 & 0 \\ 0& 0& 0 & -1
        \end{array}\right).
\end{equation}
Let us fix again the total angular momentum and decompose each subspace into Witten parity eigenspaces by
\begin{equation}\label{HDjms}
  {\cal H}^{(\pm)}_{j,m_j}= \bigoplus_{s=\pm 1}{\cal H}^{(\pm)}_{j,m_j,s}\qquad\mbox{with} \qquad {\cal H}^{(\pm)}_{j,m_j,s}=L^2(\mathbb{R}^+,r^2\rmd r)
\end{equation}
Then SUSY transformations induced by $D$ provide mappings between these eigen\-spaces
\begin{equation}\label{DcalHmjs}
  D:{\cal H}^{(\pm)}_{j,m_j,s} \to {\cal H}^{(\mp)}_{j,m_j,-s}
\end{equation}
with four radial partner Hamiltonians on $ L^2(\mathbb{R}^+,r^2\rmd r)$ given by
\begin{equation}\label{4Hpm}
  \left. H^{(\pm)}\right|_{{\cal H}_{j,m_j,s}} = \frac{P^2_r}{2m}+\frac{\ell(\ell\mp 1)}{2mr^2}+\frac{{U'}^2(r)}{2m} \mp \frac{{U''}(r)}{2m}\mp\frac{s\ell}{m\,r}\,U'(r)\,.
\end{equation}
Obviously the SUSY transformations \eqref{DcalHmjs} mixes states between the Hilbert subspaces ${\cal H}^{(+)}$ and ${\cal H}^{(-)}$. Hence, we introduce new subspaces which are invariant under SUSY transformations. These are obviously given by
\begin{equation}\label{H12}
  {\cal H}^{(1)} :=\left( \begin{array}{cc} {\cal H}^{(+)}_{j,m_j,1} & 0 \\ 0 & {\cal H}^{(-)}_{j,m_j,-1}\end{array}\right)\,,\qquad
  {\cal H}^{(2)} :=\left( \begin{array}{cc} {\cal H}^{(-)}_{j,m_j,1} & 0 \\ 0 & {\cal H}^{(+)}_{j,m_j,-1}\end{array}\right)\,,
\end{equation}
where in both subspaces the upper component belongs to $s=1$ and the lower one to $s=-1$, respectively. That is, in both subspaces we have the same matrix representation for the Witten parity and the spin-orbit operator,
\begin{equation}\label{WKinH12}
  W=\left(\begin{array}{cc} 1 & 0 \\ 0 & -1\end{array}\right)\,,\qquad K=\left(\begin{array}{cc} \ell & 0 \\ 0 & -\ell\end{array}\right)\,.
\end{equation}
The corresponding Hamiltonians read
\begin{equation}\label{H12pm}
  H^{(1)}= \left(\begin{array}{cc} H^{(1)}_+ & 0 \\ 0 & H^{(1)}_- \end{array}\right)\,,\qquad
  H^{(2)}= \left(\begin{array}{cc} H^{(2)}_+ & 0 \\ 0 & H^{(2)}_- \end{array}\right)\,,
\end{equation}
where
\begin{equation}\label{H12pm2}
\begin{array}{l}
  \displaystyle
   H^{(1)}_\pm := \frac{P^2_r}{2m}+\frac{\ell(\ell\mp 1)}{2mr^2}+\frac{{U'}^2(r)}{2m} \mp \frac{{U''}(r)}{2m}-\frac{\ell}{m\,r}\,U'(r)\,,\\[4mm]
  \displaystyle
   H^{(2)}_\pm := \frac{P^2_r}{2m}+\frac{\ell(\ell\mp 1)}{2mr^2}+\frac{{U'}^2(r)}{2m} \pm \frac{{U''}(r)}{2m}+\frac{\ell}{m\,r}\,U'(r)\,.
\end{array}
\end{equation}
That is, these pairs of SUSY Hamiltonians differ by an overall sign in $U$ and imply that if SUSY is unbroken, say for $H^{(1)}$, then SUSY is broken for $H^{(2)}$ and vice versa. Being a bit more explicit let us define the corresponding SUSY charges
\begin{equation}\label{A12}
\begin{array}{l}
  Q^{(1)}=\left(\begin{array}{cc}0 & A_1\\ A_1^\dag & 0 \end{array}\right)\qquad\mbox{with}\qquad\displaystyle
  A_1:= \frac{1}{\sqrt{2m}}\left( P_r -\rmi \frac{\ell}{r}+\rmi U'(r)\right)\,,\\[4mm]
  Q^{(2)}=\left(\begin{array}{cc}0 & A^\dag_2\\ A_2 & 0 \end{array}\right)\qquad\mbox{with}\qquad\displaystyle
  A_2:= \frac{1}{\sqrt{2m}}\left( P_r +\rmi \frac{\ell}{r}+\rmi U'(r)\right)\,,
\end{array}
\end{equation}
and observe that
\begin{equation}\label{H12pm3}
  H^{(1)}_+ =A_1A^\dag_1\,,\quad H^{(1)}_- =A^\dag_1A_1\,,\qquad
  H^{(2)}_+ =A^\dag_2 A_2\,,\quad H^{(2)}_- =A_2 A^\dag_2\,.
\end{equation}
Looking at the kernels of \eqref{A12} and their adjoints, we find
\begin{equation}\label{psizero}
  \begin{array}{lll}
    A_1\psi^{(-)}_1 =0  &\qquad\Longrightarrow\qquad & \psi^{(-)}_1(r)\sim r^{-\ell-1}\rme^{U(r)}\,, \\[2mm]
    A^\dag_1\psi^{(+)}_1 =0  &\qquad\Longrightarrow\qquad & \psi^{(+)}_1(r)\sim r^{\ell-1}\rme^{-U(r)}\, ,\\[2mm]
    A_2\psi^{(-)}_2 =0  &\qquad\Longrightarrow\qquad & \psi^{(-)}_1(r)\sim r^{\ell-1}\rme^{U(r)} \,,\\[2mm]
    A^\dag_2\psi^{(+)}_2 =0  &\qquad\Longrightarrow\qquad & \psi^{(+)}_1(r)\sim r^{-\ell-1}\rme^{-U(r)}\,.
  \end{array}
\end{equation}
Assuming that the superpotential $U$ remains well-behaved  at the origin and divergences for large $r$ like  $U(r)\to+\infty$, we realise that the SUSY system $H^{(1)}$ characterises an unbroken SUSY with ${\rm ker\,} A^\dag_1\neq 0$ and ${\rm ker\,} A_1 = 0$, whereas $H^{(2)}$ has broken SUSY as ${\rm ker\,} A_2 = 0= {\rm ker\,} A_2^\dag$. The sole zero-energy ground state is then given by $\psi_1^{(+)}$ in \eqref{psizero}.

We conclude that for rather generic superpotentials we arrive at two SUSY systems, the one in ${\cal H}^{(1)}$ with an unbroken SUSY phase and the one in ${\cal H}^{(2)}$ with broken SUSY. All such systems will have a generic spin-orbit interacting of the form $\pm\frac{\ell U'(r)}{mr}$. We note that these two phases can be interchanged when changing the sign of the superpotential $U$. In general such a change is achieved when changing sign of one or several parameters characterising the superpotential. This switching between unbroken and broken SUSY phase has been observed before in various one-dimensional Witten models \cite{Dutt1993,Asim2001}. Let us now discuss a few examples in detail.

%Matrix representation of Q_c

\subsection{Example 1: The Pauli oscillator}
In analogy to the Dirac oscillator let us consider the case of a quadratic superpotential
\begin{equation}\label{UPauliOsc}
  U(r):= \frac{m}{2}\,\omega\, r^2\,,\qquad \omega > 0\,.
\end{equation}
Following above discussion it is clear that the subsystem characterised by $H^{(1)}$ corresponds to the unbroken SUSY phase with $H^{(1)}_\pm = \frac{P^2_r}{2m}+V^{(1)}_\pm$, where
\begin{equation}\label{V1pmPO}
  V^{(1)}_\pm(r) = \frac{\ell(\ell\mp 1)}{2mr^2} +\frac{m}{2}\omega^2 r^2 -   \omega\left(\ell \pm \frac{1}{2}\right)\,,\qquad
  A_1=\frac{1}{\sqrt{2m}}\left[P_r -\rmi\left(\frac{\ell}{r}-m\omega r\right)\right]\,.
\end{equation}
The corresponding eigenvalues of $H^{(1)}_\pm$ are given by $E^{(1)}_n= 2n\omega$, where $n\in\mathbb{N}_0$ for $H^{(1)}_+$ and $n\in\mathbb{N}$ for $H^{(1)}_-$. That is the zero-energy SUSY ground state belongs to $H^{(1)}_+$ as expected by the general discussion above.

The subsystem belonging to the SUSY pair  $H^{(2)}_\pm = \frac{P^2_r}{2m}+V^{(2)}_\pm$ is in the broken SUSY phase with
\begin{equation}\label{V2pmPO}
  V^{(2)}_\pm(r) = \frac{\ell(\ell\mp 1)}{2mr^2} +\frac{m}{2}\omega^2 r^2 +  \omega\left(\ell \pm \frac{1}{2}\right)\,,\qquad
  A_2=\frac{1}{\sqrt{2m}}\left[P_r +\rmi\left(\frac{\ell}{r}+m\omega r\right)\right]\,.
\end{equation}
Here both partner Hamiltonians have identical spectrum given by $E^{(2)}_n=\omega(2n+2\ell+1)$ with $n\in\mathbb{N}_0$.

Let us remark that this is not a new result. In fact, the Witten model for the radial harmonic oscillator is known to exhibit both, broken and unbroken SUSY realisations \cite{Junker1996}. Here however, we derived both in a common setting within a fixed total angular momentum scheme based on the spin-$\frac{1}{2}$ Pauli Hamiltonian. Both potential pairs, \eqref{V1pmPO} and \eqref{V2pmPO}, are known to be shape invariant and allow for an exact solution of the eigenvalue problem.

\subsection{Example 2: A linear superpotential}
As a second example we consider a linear superpotential of the form
\begin{equation}\label{Ulinear}
  U(r):= \gamma r\,,\qquad \gamma > 0\,.
\end{equation}
Again subsystem ${\cal H}^{(1)}$ will be in the unbroken and system ${\cal H}^{(2)}$ in the broken SUSY phase as $\lim_{r\to\infty}U(r)=+\infty$. The corresponding partner potentials read
\begin{equation}\label{V12pmlinear}
  \displaystyle
   V^{(1)}_\pm(r)=\frac{\ell(\ell\mp 1)}{2mr^2}+\frac{\gamma^2}{2m}-\frac{\ell\gamma}{m}\,\frac{1}{r}\,,\qquad
   V^{(2)}_\pm(r)=\frac{\ell(\ell\mp 1)}{2mr^2}+\frac{\gamma^2}{2m}+\frac{\ell\gamma}{m}\,\frac{1}{r}\,.
\end{equation}
We arrive at Coulomb like potentials with an angular momentum dependent coupling constant being attractive in the unbroken SUSY phase and repulsive in the broken SUSY phase. Note that we basically arrive at the Coulomb problem as discussed in subsection \ref{subsec52}, however, with $\alpha =\frac{\ell\gamma}{m}$ and $\alpha =-\frac{\ell\gamma}{m}$, respectively. Only in the unbroken SUSY case we do have a discrete spectrum given by
\begin{equation}\label{Enlinear}
  E^{(1)}_n = \frac{\gamma^2}{2m}\left[1-\left(\frac{\ell}{n+\ell}\right)^2\right]
\end{equation}
with $n\in\mathbb{N}_0$ for $H^{(1)}_+$ and $n\in\mathbb{N}$ for $H^{(1)}_-$. This way we have yet arrived at another exactly solvable radial Coulomb-like problem. Here we note that this exactly solvable quantum model can be associate with a shape invariance which consists of both a translation and scaling of the potential parameters. Let us denote these parameters explicitly by writing $V^{(1)}_\pm(r)\equiv V^{(1)}_\pm(\ell,\gamma;r)$. Then we can establish the shape-invariance relation
\begin{equation}\label{V1pmSI}
  V^{(1)}_-(\ell_{n},\gamma_{n};r) = V^{(1)}_+(\ell_{n+1},\gamma_{n+1};r) + \frac{1}{2m}\left(\gamma_n^2-\gamma^2_{n+1}\right)
\end{equation}
with the parameter translation and scaling relation given by
\begin{equation}\label{V1transcale}
    \ell_{n+1}=\ell_n + 1 = \ell_0 +n \,,\qquad \gamma_{n+1}=\gamma_n\frac{\ell_n}{\ell_{n+1}}= \gamma_0\frac{\ell_0}{\ell_0+n}\,.
\end{equation}
With $\ell_0=\ell$ and $\gamma_0=\gamma$ the shape-invariance relation \eqref{V1pmSI} reproduces the spectrum \eqref{Enlinear}.  The corresponding eigenfunction may be obtain from the ground state
\begin{equation}\label{psi1linear}
  \psi^{(+)}_1(\ell_n, \gamma_n;r)=\frac{2\gamma_n}{\sqrt{2\ell_n}!}\,r^{\ell_n-1}\rme^{-\gamma_n r}
\end{equation}
by a successive application of operator $A_1$with corresponding parameters.

\subsection{Example 3: A generic power-law superpotential}
\begin{figure}[t]
\includegraphics[scale=1]{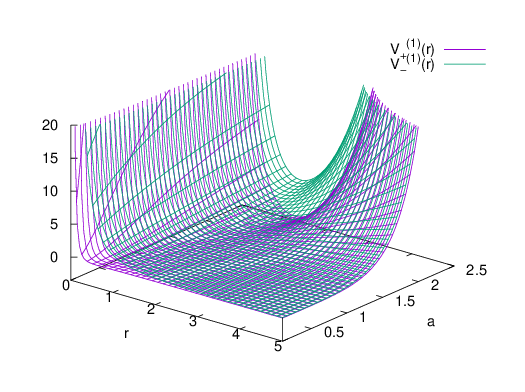}
\includegraphics[scale=1]{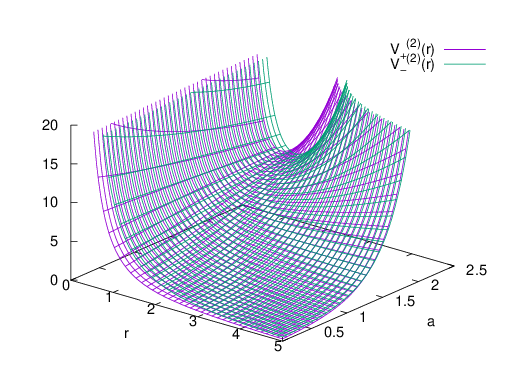}
\caption{The partner potentials \eqref{V12a} for $2m=1=\gamma$ and $\ell=2$. The left graph shows the pair $ V_\pm^{(1)}$ associated with unbroken SUSY and the right graph presents the broken SUSY pair $ V_\pm^{(2)}$.}
\end{figure}
Both examples discussed above belong to the family of power-law superpotential indexed by a parameter $a>0$ and defined via
\begin{equation}\label{Upowerlaw}
  U(r):= \frac{\gamma}{a}\,r^a\,,\qquad \gamma > 0\,.
\end{equation}
Obviously, our example 1, the Pauli oscillator, is the family member with $a=2$  and $\gamma=m\omega$. Our second example is the member with $a=1$. Let us now consider the case of an arbitrary $a>0$.
The pair of partner potentials $V^{(1)}_\pm$ and $V^{(2)}_\pm$, belong to the unbroken and broken SUSY case, respectively,  are given by
\begin{equation}\label{V12a}
\begin{array}{l}
  \displaystyle
  V_\pm^{(1)}(r)=\frac{1}{2m}\left[\frac{\ell(\ell\mp 1)}{r^2}+\gamma^2 r^{2a-2}-2\gamma\ell\left(1\pm\frac{a-1}{2\ell}\right)r^{a-2}\right]\,,\\[4mm]
  \displaystyle
  V_\pm^{(2)}(r)=\frac{1}{2m}\left[\frac{\ell(\ell\mp 1)}{r^2}+\gamma^2 r^{2a-2}+2\gamma\ell\left(1\pm\frac{a-1}{2\ell}\right)r^{a-2}\right]\,.
\end{array}
\end{equation}
Graphs of both pairs of partner potentials are shown in figure 1 for fixed parameters $2m=1=\gamma$, $\ell= 2$ and a range of the power exponent $0<a<2.5$ including the examples 1 and 2 of the Pauli oscillator and the linear superpotential.
The SUSY ground state belonging to the zero energy eigenvalue of $H^{(1)}_+$ is given via \eqref{psizero} and explicitly reads
\begin{equation}\label{psizeroa}
  \psi_1^{(+)}(r) = N\,r^{\ell-1}\rme^{-(\gamma/a) r^a}\qquad\mbox{with}\qquad N^{-2}:=\frac{1}{2\gamma}\left(\frac{a}{2\gamma}\right)^{\frac{2\ell+1}{a}-1}\Gamma\left(\frac{2\ell+1}{a}\right)\,,
\end{equation}
which is properly normalized with respect to the measure $\rmd r\,r^2$.

Let us discuss the limiting case $a\to 0$. Despite our requirement $a>0$ we can put $a=0$ in the two pairs \eqref{V12a}, which corresponds to a logarithmic superpotential $U(r)=\gamma\ln r$ resulting in
\begin{equation}\label{V120}
V_\pm^{(1)}(r)=\frac{1}{2mr^2}(\ell -\gamma)(\ell -\gamma \mp 1)\,,\qquad
  V_\pm^{(2)}(r)=\frac{1}{2mr^2}(\ell +\gamma)(\ell +\gamma \mp 1)\,.
\end{equation}
Hence, we are basically back to the free particle case discussed in section \ref{subsec51} with a rescaled angular momentum $\ell\to \ell-\gamma$ and $\ell\to \ell+\gamma$, respectively. Note that both pairs in \eqref{V120} are shape invariant under the translation $\ell\to\ell+1$ with $ V_\pm^{(i)}\to  V_\mp^{(i)}$.

\begin{figure}[t].
\includegraphics[scale=1]{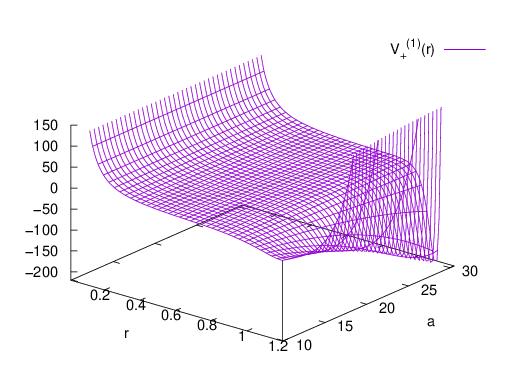}
\includegraphics[scale=1]{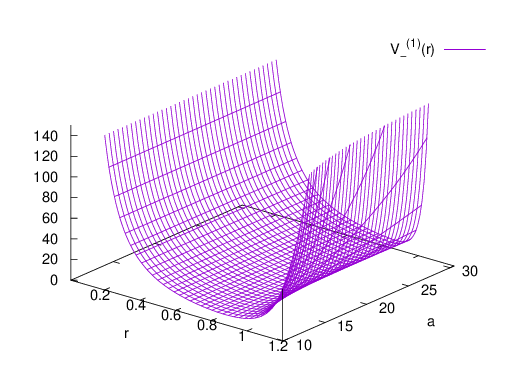}
\caption{The partner potentials $ V_\pm^{(1)}$ for $2m=1=\gamma$, $\ell=2$ and large values of the exponent $a\in[10,30 ]$. The left graph shows $ V_+^{(1)}$ indicating that the limit $a\to\infty$
simulates a Neumann boundary condition at $r=1$. The left graph shows $ V_-^{(1)}$ indicating a Dirichlet boundary condition at $r=1$.}
\end{figure}
\begin{figure}[t].
\includegraphics[scale=1]{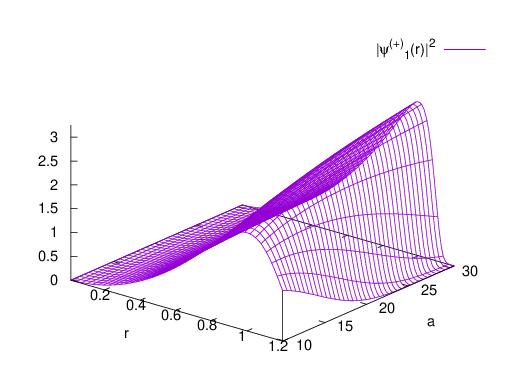}
\caption{The probability density corresponding to the zero-energy ground state \eqref{psizeroa} for the same parameters as in figure 2. We clearly see a localisation at $r=1$ for large $a$ as expected.}
\end{figure}
More interesting is the case of large $a\to\infty$. Let us look at the $\gamma$-dependent part of the two pairs of potential \eqref{V12a} by defining
\begin{equation}\label{Wpm}
  W_\pm(r):= \gamma^2r^{2a-2}\pm a\gamma r^{a-2}\,.
\end{equation}
Obviously, in the limit for large $a$ we have the relation
\begin{equation}\label{V12alarge}
   V_\pm^{(1)}(r)\approx\gamma^2r^{2a-2} + W_\pm(r) \approx V_\mp^{(2)}(r)\,,\qquad a\gg 1\,.
\end{equation}
In addition, we observe that \eqref{Wpm} simulates a spherical potential well of unit radius when taking the limit $a\to\infty$,
\begin{equation}\label{Walarge}
  \lim_{a\to\infty} W_\pm(r)=\left\{\begin{array}{cc}0 \qquad & 0<r<1\\[2mm] +\infty & r>1 \end{array}\right.\,.
\end{equation}
Hence, in this limit  both, $V_\pm^{(1)}$ and $V_\pm^{(2)}$ basically represent the effective radial potential for free particle in a box of unit radius.

In figure 2 we have plotted $V_\pm^{(1)}$  for the same parameters as in figure 1, the only difference being the range for $a$, namely $10<a<30$. Here we see for $V_+^{(1)}$ a narrow deep trough developing near $r \lessapprox 1$, which simulates a van Neumann boundary condition at $r=1$ in the limit $r\to\infty$. However, for $V_-^{(1)}$ we see that for increasing $a$ the development of a repulsive wall simulating a Dirchlet boundary condition at $r=1$. A similar behaviour was observed in the one-dimensional Witten model with a power-law SUSY potential \cite{Junker1996}. Despite the fact that $V^{(1)}_\pm(r)\simeq V^{(2)}_\mp(r)$ for large $a$, the unbroken SUSY phase has an additional zero-energy ground state \eqref{psizeroa}. The corresponding probability density $|\psi_1^{(+)}(r)|^2$ is shown in figure 3 for same parameters as in figure 2. We observe that for large $a$ this state becomes more and more localised near $r=1$.

In conclusion we can state that the generic power-law superpotential provides an approximate model for a particle in a spherical box with both, von Neumann and Dirichlet boundary conditions, when taking the limit of large powers $a\gg 1$. For any finite $a$ we observe the previously mentioned switch between broken and unbroken SUSY phase when changing sign $\gamma\to -\gamma$.

\section{Summary and Outlook}\label{sec7}
In this work we have studied the SUSY structure of spherically symmetric Pauli Hamiltonians. Due to its rotationally invariance the spin-orbit operator $K$ is a conserved quantity and, when restricting the system to a fixed total angular momentum subspace characterised by fixed quantum number $j$, the eigenvalues of $K$ can take on the two values $\pm(j+\frac{1}{2})$ and may be used to introduce a grating of this subspace. That is, it takes over the role of the Witten operator. One then only needs to choose a proper vector operator $\vec{v}$, which than allows to construct a supercharge and in turn a SUSY Hamiltonian via \eqref{SUSYradial}. We have considered the cases of a pure scalar (I) and a pure vector potential (II).

For case (I) our choice $\vec{v}=\vec{P}$ resulted in a pair of radial free particles with orbital angular momentum $\ell -1 $ and $\ell$ forming a SUSY system. The SUSY transformations turned out to reproduce the well-known recursion relations of the spherical Bessel functions. As second example we took the Laplace-Runge-Lenz vector for the construction of the supercharge resulting in the SUSY Hamiltonian \eqref{HSUSYCoulomb}, which is basically that of the non-relativistic Coulomb problem. Restricted on a subspace with fix $j$ where $K^2=\ell^2$ is constant, we rediscovered the radial Witten model for the Coulomb problem. We note here that the radial harmonic oscillator, which also exhibits a SUSY structure, cannot be accommodated into this framework and let us to consider the second case.

In case (II), inspired by the Dirac oscillator approach, we considered a purely imaginary radially symmetric vector potential which is represented by the gradient of a radially symmetric function call superpotential. In this case the supercharge becomes complex and the general formalism of section 4 forced us to double our Hilbert space in order to arrive at some nontrivial results. The pair of SUSY Hamiltonians \eqref{HpmD} then characterises a non-relativistic quantum system with an external potential depending on derivatives of that superpotential and differ by the sign of the
superpotential. Rearranging these pairs results in two subsystems, one with broken and one with unbroken SUSY for suitable superpotentials. Several examples like the Pauli oscillator, a Coulomb like system with spin-orbit coupling  and the generic power-law superpotential are discussed in detail. Whereas for the Pauli oscillator we obtained the well-known Witten model of the radial harmonic oscillator being translational shape invariant, the linear superpotential resulting in a Coulomb like radial system provided us with a shape invariance which requires both, a translation in the angular momentum quantum number and a scaling in the potential parameter. The translation of parameters in shape invariance is well known \cite{Junker1996} and also a multiplication of the parameters has been utilised in shape invariance \cite{Barclay1993,Rizek1996}. Here for the first time we have found a system which requires both types of parameter transformations.

\backmatter

%\pagebreak

%\bmhead{Acknowledgements}
%Acknowledgements are not compulsory. Where included they should be brief. Grant or contribution numbers may be acknowledged.
%Please refer to Journal-level guidance for any specific requirements.
\section*{Declarations}
\begin{itemize}
\item The author received no financial support for the research, authorship, and/or publication of this article.
\item The author has no conflicts of interest to disclose.
\item The author declares that no data had been used nor generated supporting the findings of this study.
\end{itemize}

%%===================================================%%
%% For presentation purpose, we have included        %%
%% \bigskip command. Please ignore this.             %%
%%===================================================%%

%%===========================================================================================%%
%% If you are submitting to one of the Nature Portfolio journals, using the eJP submission   %%
%% system, please include the references within the manuscript file itself. You may do this  %%
%% by copying the reference list from your .bbl file, paste it into the main manuscript .tex %%
%% file, and delete the associated \verb+\bibliography+ commands.                            %%
%%===========================================================================================%%

%\bibliography{sn-bibliography}% common bib file

\end{document}